\documentclass[11pt]{article}
\usepackage[a4paper, lmargin=25mm, rmargin=25mm, tmargin=15mm, bmargin=28mm]{geometry}

\usepackage{algorithm}
\usepackage{algpseudocode}
\usepackage{amsfonts}
\usepackage{amsmath}
\usepackage{amsopn}
\usepackage{amssymb}
\usepackage{array}
\usepackage[affil-it]{authblk}
\usepackage[english]{babel}
\usepackage{bbold}
\usepackage{blindtext}
\usepackage{booktabs}
\usepackage{caption}
\usepackage{changepage}
\usepackage{fancyhdr}
\usepackage{geometry}
\usepackage{graphicx} 
\usepackage{ifthen}
\usepackage{mathptmx}  
\usepackage{multirow}
\usepackage[numbers]{natbib}
\usepackage{soul}
\usepackage{subcaption}
\usepackage{tikz}
\usepackage[explicit]{titlesec}
\usepackage{url}
\usepackage{xcolor}

\newcommand{\mvec}[1]{\boldsymbol{#1}}

\title{
\vspace{10mm}
\fontsize{14}{14}\selectfont\textbf{UNCERTAINTY QUANTIFICATION\\ ON SPENT NUCLEAR FUEL WITH LMC}
}
\author[1,2]{\fontsize{11}{11}\selectfont\textbf{Arnau Alb\`a}}
\author[1]{\fontsize{11}{11}\selectfont\textbf{Andreas Adelmann}}
\author[1]{\fontsize{11}{11}\selectfont\textbf{Dimitri Rochman}}
\affil[1]{Paul Scherrer Institut, Forschungstrasse 111, 5232 Villigen, Switzerland}
\affil[2]{ETH Zurich, R\"amistrasse 101, 8092 Zurich
 Switzerland}
\affil[*]{arnau.albajacas@psi.ch}
\date{}

\providecommand{\customAbstract}[1]
{
  \begin{center}
      \textbf{ABSTRACT}
  \end{center}
  \begin{adjustwidth}{10mm}{10mm}
  \textit{#1}
  \end{adjustwidth}
}

\providecommand{\keywords}[1]
{
  \begin{center}
      \textbf{KEYWORDS}
  \end{center}
  
  \begin{adjustwidth}{10mm}{10mm}
  \textit{#1}
  \end{adjustwidth}
}

\titleformat{\section}
  {\fontsize{11}{11}\selectfont\bf}{\hspace{5mm}\thesection}{0pt}{. \MakeUppercase{#1}}
\titlespacing\section{0pt}{11pt plus 4pt minus 2pt}{0pt plus 2pt minus 2pt}
\titleformat{\subsection}
  {\fontsize{11}{11}\selectfont\bf}{\hspace{5mm}\thesubsection}{0pt}{. #1}
\titlespacing\subsection{0pt}{11pt plus 4pt minus 2pt}{0pt plus 2pt minus 2pt}

\begin{document}
\parindent=0pt
\parskip=11pt

\maketitle

\customAbstract{
The recently developed method Lasso Monte Carlo (LMC) for uncertainty quantification is applied to the characterisation of spent nuclear fuel. The propagation of nuclear data uncertainties to the output of calculations is an often required procedure in nuclear computations. Commonly used methods such as Monte Carlo, linear error propagation, or surrogate modelling suffer from being computationally intensive, biased, or ill-suited for high-dimensional settings such as in the case of nuclear data. The LMC method combines multilevel Monte Carlo and machine learning to compute unbiased estimates of the uncertainty, at a lower computational cost than Monte Carlo, even in high-dimensional cases. Here LMC is applied to the calculations of decay heat, nuclide concentrations, and criticality of spent nuclear fuel placed in disposal canisters. The uncertainty quantification in this case is crucial to reduce the risks and costs of disposal of spent nuclear fuel. The results show that LMC is unbiased and has a higher accuracy than simple Monte Carlo.
}

\keywords{Uncertainty quantification, Nuclear data, Decay heat, Criticality, Spent nuclear fuel}
\thispagestyle{fancy}

\section{Introduction}
Nuclear codes make predictions about spent nuclear fuel (SNF), that are then used by regulators to make informed decisions that reduce the risks and costs of SNF storage. It is therefore important that the uncertainties of these codes are quantified and taken into account when making predictions. This work studies methods to calculate the uncertainty in quantities used to characterise SNF, namely the decay heat, nuclide inventory, and the k-effective of the SNF canisters.

One of the main sources of uncertainty in neutron transport codes is nuclear data, which include tens of thousands of uncertain parameters such as neutron cross-sections, fission product yields, neutron multiplicity, or the fission spectra. The uncertainty propagation from nuclear data has traditionally been done with either linear perturbation methods, or Monte Carlo (MC) \cite{rochman_nuclear_2016}. The former method assumes the linearity of the response function, and requires computing the sensitivity of the response with respect to each input dimension, thus it is potentially biased and its cost increases with the input dimension. On the other hand, MC methods are dimension-independent and make no (or weak) assumptions on the distribution of the output. However, MC is known to have a slow convergence and needing a large number of simulations, rendering the method computationally expensive. 

In the case of Switzerland, over 12 thousand SNF assemblies are expected \cite{solans_loading_2020}, corresponding to a 60-year operation of currently operating nuclear power plants. Computing the nuclear data uncertainty requires close to a thousand simulations per assembly, with each simulation running in an order of ten hours on a single CPU core. This amounts to a rough estimate of at least 100 million expected CPU hours for all the uncertainty calculations required in Switzerland. Therefore, it is of high interest to investigate methods that reduce the computational costs of uncertainty quantification (UQ) from nuclear data.

Recent advances in machine learning have shown large increases in speed in several domains, with the use of fast surrogate models that replace the computationally expensive codes (e.g. \cite{ebiwonjumi_machine_2021}). However, surrogate models, similarly to linear perturbation methods, introduce model bias by making assumptions on the response function, and suffer from the curse of dimensionality \cite{beyer_when_1999}: since nuclear data has tens of thousands of dimensions, training a surrogate model would require tens of thousands of expensive simulations. Other UQ approaches that focus on dimensionality reduction, such as principle component analysis \cite{jolliffe_principal_2016} or active subspaces \cite{subspace_2014}, have not been investigated in this work. However, these methods are in general also biased as they ignore the effect of certain inputs dimensions.

In this work, the Lasso Monte Carlo (LMC) method \cite{alba2023lasso} is applied to UQ from nuclear data. This method uses a Lasso \cite{tibshirani_regression_1996} surrogate model, which is a simple sparse linear model that requires few training samples, and then computes the moments of the response function with Multilevel Monte Carlo \cite{giles_multilevel_2008}, which corrects for the bias of the surrogate model. Overall, the method is unbiased and more accurate than MC. This is computationally verified with UQ calculations on a variety of quantities for two fuel assembly (FA) examples, one uranium oxide case (UO2) and one mixed oxide case (MOX). Note that this paper only includes a brief explanation of the LMC method without any proofs, as the full details of the method can be found in \cite{alba2023lasso}.

The rest of the paper is organised as follows. In the \textit{methodology section} several UQ methods are explained, with emphasis on the most recent method, LMC. This section also includes details regarding the computational tools, the uncertainties in nuclear data, and the two studied fuel assemblies. The results section compares LMC to the other described methods, all of which are applied to a range of quantities of interest. Finally, the conclusion summarises the work and discusses future research directions.

\section{Methodology}
\subsection{Nuclear Codes}
The burnup calculations were carried out with the nuclear lattice code CASMO5 \cite{rhodes_casmo-5_2006}. A single fuel assembly was modelled as a 2D slice with reflecting boundaries. Starting from the fresh fuel characteristics, CASMO5 repeatedly solves the decay equation and the nuclear transport equation to simulate all the burnup and cooling cycles of the assembly, until its end of life (EOL). The output quantities, that were of interest in this work, were the decay heat w.r.t. cooling time, and the nuclide inventory at EOL.

For the UO2 simulations, large burnup steps with cycle-wise averaged quantities were used (as was done in \cite{shama_validation_2022}), and a quarter-symmetry was considered. These simulations needed approximately 10 minutes each. For the MOX case, a detailed burnup history with in-cycle variations was used, since it was available, and the full 2D slice of the FA was simulated, without any symmetry assumptions. The MOX burnup simulations needed approximately 3 hours each. All simulations and further calculations were carried out on single CPU cores, specifically on Intel Xeon Gold 6152 processors with 2.10 GHz and 4 GB of RAM.

The PSI in-house code Shark-X \cite{wieselquist_psi_2013, aures_benchmarking_2017} was used to perturb the nuclear data in the 20 energy groups used by CASMO5, based on the covariances of ENDF/B-VII.1 \cite{chadwick_endfb-vii1_2011}. The perturbed quantities were the neutron elastic and inelastic cross-sections, neutron capture cross-sections, fission cross-sections, fission product yields, fission spectra, and neutron multiplicity. The total number of perturbed nuclear data values was 15 557.

\begin{figure}[H]
    \centering
    \begin{tikzpicture}
        \node[anchor=south west,inner sep=0] at (0,0) {\includegraphics[width=0.5\textwidth, trim={30 20 0 0},clip]{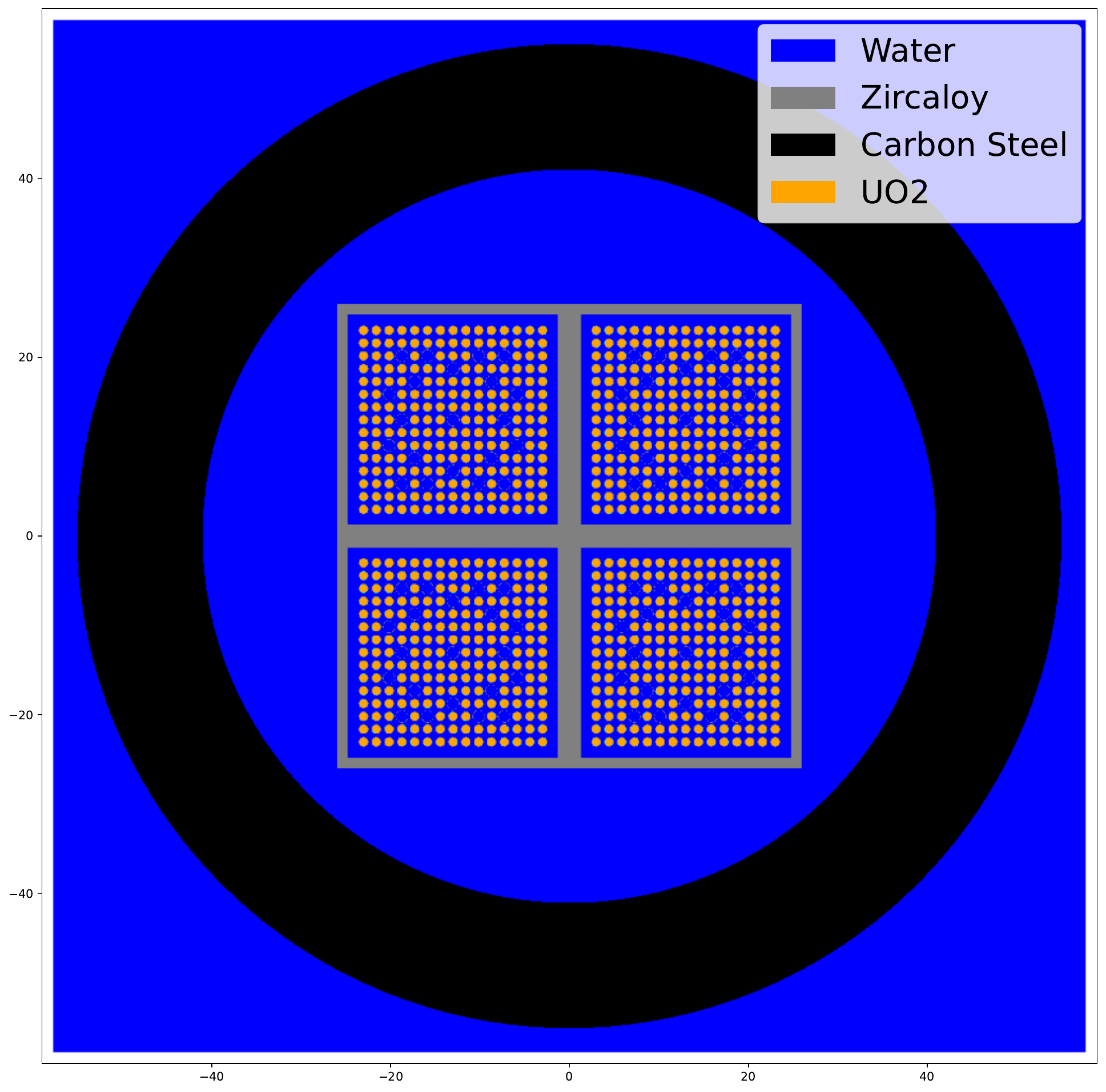}};
        \draw[red, line width=3.5pt, rounded corners] (0,7.8) node[anchor=north]{}
        -- (4, 3.9) node[anchor=north]{}
        -- (4,7.8) node[anchor=south]{}
        -- cycle;
    \end{tikzpicture}
    \caption{Canister design employed in the OpenMC criticality calculations, with shape based on the design from \cite{vasiliev_preliminary_2019}. The inner and outer radii of the canister are 41 cm and 55 cm respectively, and the four SNF assemblies are identical. To speedup calculations, an 8-fold symmetry simulation was assumed, hence only the part in the red triangle was modelled, with reflecting boundary conditions.}
    \label{fig:canister}
\end{figure}

The criticality calculations were preformed with the open source Monte Carlo code OpenMC \cite{romano_openmc_2015}. These calculations were only carried out for the UO2 assemblies. The modelled system was an SNF canister containing 4 identical depleted FAs, surrounded by a 14 cm layer of carbon steel, and filled with water. Fig \ref{fig:canister} shows the model  of a spent fuel canister, based on the dimensions and arrangement used in \cite{vasiliev_preliminary_2019} (although here the FAs are placed in zircaloy boxes rather than carbon steel boxes as in the original work). The calculations were done in 2D, and with an 8-fold symmetry and reflecting boundaries. The nuclide inventory of the depleted FAs was obtained from the CASMO5 output files, hence one criticality calculation was carried out for each burnup simulation. The nuclear data employed by OpenMC was ENDF/B-VII.1 \cite{chadwick_endfb-vii1_2011} (formatted ACE files distributed by OpenMC developers), and was used without perturbations. Each OpenMC simulation required approximately 6 hours, and was carried out with $2\cdot 10^5$ neutron histories for 400 batches (of which the first 100 were inactive), to obtain a statistical uncertainty $\sigma_{stat}=10$ pcm. For the criticality calculations a RAM of 32 GB was used.

\subsection{Monte Carlo Uncertainty Quantification}
Consider a quantity of interest $q$ that is given as an output of a calculation $q=f(\mvec{x})$, where $\mvec{x}\in\mathbb{R}^d$ is a vector containing the $d$ input parameters. Consider now that the input is a random variable $X$ with a known mean and covariance matrix, from which input vectors can be sampled $x_1, x_2,...$ . Then the output is also a random variable $Q=f(X)$, and one wants to calculate the mean $\mathbb{E}[Q]$ and standard deviation $\sqrt{\text{Var}[Q]}$ (i.e. the uncertainty) of the quantity of interest. In the case of this paper $q$ could be the decay heat, nuclide concentrations, or k-effective, the calculation $f$ is a burnup computation with CASMO5, or CASMO5 followed by a criticality calculation with OpenMC, and the vector $x$ contains $d=15\,557$ nuclear data parameters.

With simple Monte Carlo (MC), a set of input vectors $x_1, x_2, ..., x_N$ is sampled and used as input for $N$ calculations. Then the sample mean and variance can be estimated from the outputs $q_1, q_2, ..., q_N$ with the usual sample estimators (eq. \ref{eq:MC_estimators}). With this method, the expected errors on $\mu_N$ and $\sigma_N$ converge to zero with the number of samples as $\mathcal{O}(1/\sqrt{N})$, i.e. the method is unbiased and dimensionality independent. The disadvantage of MC is that the convergence is slow, and thus $f$  needs to be executed a large number of times to obtain a small error in the estimations. As an example, in this work it was found that approximately $N=1000$ was required to obtain a $1\%$ error in the estimator $\sigma_N$ for the decay heat.

\begin{equation}
    \mu_N = \frac{1}{N}\sum_{i=1}^Nq_i,\quad\sigma^2_N = \frac{1}{N-1}\sum_{i=1}^N \left(q_i - \sum_{j=1}^N \frac{q_j}{N}\right)^2\,.
    \label{eq:MC_estimators}
\end{equation}

\subsection{Lasso Monte Carlo (LMC)}
\label{sec:LMC}
The LMC method, recently introduced in \cite{alba2023lasso}, uses a combination of machine learning (ML) and multilevel Monte Carlo (MLMC), to estimate the mean and variance of $Q$. The following paragraphs briefly explain how ML and MLMC can each be used for UQ, before finally introducing the LMC method and how it combines these two techniques. All of the following methods use the same data as simple MC, i.e. the $N$ input-output samples $x_1,x_2,...,x_N$ and $q_1,q_2,...,q_N$, and an additional set of $M$ inputs $z_1,z_2,...,z_M$, where it is assumed that $M\gg N$.

In recent years there has been a trend in the use of  ML to reduce computational costs in methods that use repeated calculations. With this approach, the set of $N$ input-output pairs is used to train a fast surrogate model $\widetilde f$. Then, $\widetilde f$ is used for the other $M$ evaluations to obtain the outputs $\widetilde q_1, \widetilde q_2,...,\widetilde q_M$, which are used for the usual estimators (eq. \ref{eq:MC_estimators}). The bottleneck of this approach is evaluating $f$ $N$ times, since training the surrogate model and evaluating it $M$ times has a negligible cost. Therefore, this method is clearly only advantageous if $N<M$. However, since ML techniques suffer from the curse of dimensionality, the required size of the training set increases exponentially with the input dimension $d$, and any model trained with $N<d$ will be heavily biased, and produce biased estimates of the mean and variance. In the case of nuclear data, the input dimension is $d>10^4$, hence using $N=1000$ as in the simple MC case will lead to poor results. In conclusion, using ML for UQ, when the input dimension is large, has a higher computational cost than simple MC.

In the remaining part of the paper it will be assumed that $\widetilde f$ is a Lasso surrogate model \cite{tibshirani_regression_1996}. While it is possible, in principle, to use any other ML technique for the methods explained here, Lasso is a natural choice in cases where the training set is small $N\ll d$, which is the case in all the examples in this paper. The Lasso is a simple sparse linear model, i.e. it fits a model of type $\widetilde f(x)=\beta\cdot x + b$, where the weight vector $\beta$ is sparse. The method of fitting the model (coordinate descent) ensures that, the weights of $\beta$ corresponding to input variables with small correlations to the output, are set to zero.

The MLMC \cite{giles_multilevel_2008, krumscheid_quantifying_2020} approach is similar to using ML, but it produces unbiased results. Here, the $N$ input-output samples are split into a training set of size $N_{tr}$ and an evaluation set of size $N_{eval}$, with $N_{tr} + N_{eval} = N$. The training set is again used to train a surrogate $\widetilde f$, which is used to evaluate the $M$ and $N_{eval}$ samples. Then, all of the evaluations of $f$ and $\widetilde f$ are used together in the \textit{multilevel estimators} (eq. \ref{eq:MLMC_estimators}). These estimators are unbiased and converge at a rate $\mathcal{O}\left(1/\sqrt{N_{eval}}\right)$. Despite MLMC being unbiased, there is no guarantee that for a given sample size $N$ it will be more or less accurate than simple MC, and its accuracy largely depends on the choice of $N_{tr}$ and $N_{eval}$.

\begin{equation}
\begin{split}
    \mu_{N_{eval},M} &= \frac{1}{M}\sum_{i=1}^{M} \widetilde f(\mvec z_i) + \frac{1}{N_{eval}}\sum_{i=1}^{N_{eval}} \left[f(\mvec x_i) - \widetilde f(\mvec x_i)\right] \,,\\
    \sigma^2_{N_{eval},M} &= \frac{1}{M-1}\sum_{i=1}^M \left(\widetilde f(z_i) - \sum_{j=1}^M \frac{\widetilde f(z_j)}{M}\right)^2 \\
    &+ \frac{1}{N_{eval}-1}\sum_{i=1}^{N_{eval}}\left[ \left(f(x_i) - \sum_{j=1}^{N_{eval}} \frac{f(x_j)}{N_{eval}}\right)^2 - \left(\widetilde f(x_i) - \sum_{j=1}^{N_{eval}} \frac{\widetilde f(x_j)}{N_{eval}}\right)^2 \right]\,.
    \label{eq:MLMC_estimators}
\end{split}
\end{equation}

The LMC method can be seen as a variation of MLMC, that removes the need to optimally balance the training and evaluation sets, and guarantees that its estimates will be equally or more accurate than simple MC. Here the $N$ input-output samples are split into $S$ subsets of equal size $N/S$. Then for each subset a surrogate is trained and the multilevel estimates are calculated $\mu_{N/S,M}$ and $\sigma_{N/S,M}$ with eq. \ref{eq:MLMC_estimators}. The LMC estimators (eq. \ref{eq:final_LMC}) are simply the mean of these $S$ multilevel estimates. In the LMC paper \cite{alba2023lasso} it is shown that, under mild conditions on $f$, if the surrogate is a Lasso model and $M\gg N$, then the LMC estimators $\mathcal{M}_{N,M}$ and $\Sigma_{N,M}$ are equally or more accurate than simple MC for a given $N$. Therefore, to obtain the same error in the estimates as simple MC, LMC will require less expensive evaluations of $f$. The full LMC method is shown in Algorithm \ref{alg:LMC}, and table \ref{tab:UQ_methods} compares the four explained methods.

\begin{algorithm}
\caption{The Lasso Monte Carlo Algorithm}\label{alg:LMC}
\begin{algorithmic}[1]
\Require The training sets $\{\mvec x_1, ... \mvec x_N\}$ and $\{\mvec z_{1}, ... \mvec z_{M}\}$, and model $f:\mathbb{R}^d\mapsto \mathbb{R}$.

\State Compute $f(\mvec x_1), ..., f(\mvec x_N)$.
\State Split the $N$ input-output samples into $S$ subsets of size $\frac{N}{S}$ each. The subsets are $s_1,s_2,...,s_S$.

\For{$p=1 \ldots S$}
    \State The p-th subset is the evaluation set, containing the samples $s_p=\{x^\prime_1,x^\prime_2,...,x^\prime_{N/S}\}$.
    \State Make a training set of size $N\frac{S-1}{S}$ by joining the other subsets $s_1\cup s_2\cup...\cup s_{p-1}\cup s_{p+1}\cup...\cup s_S$.
    \State Fit a Lasso model $\widetilde f_{p}$ to the training set.
    \State Compute $\widetilde f_p(\mvec z_1),\widetilde f_p(\mvec z_2),...,\widetilde f_p(\mvec z_M)$, and $\widetilde f_p\left(x^\prime_1\right), \widetilde f_p\left(\mvec x^\prime_2\right), ..., \widetilde f_p\left(\mvec x^\prime_{N/S}\right)$ .
    \State Combine the $N/S$ evaluations of $f$ and $\widetilde f_p$ with the $M$ evaluations of $\widetilde f_p$ to compute the two-level estimators $(\mu_{N/S,M})_{p}$ and $(\sigma^2_{N/S,M})_{p}$ (eq. \ref{eq:MLMC_estimators}).
\EndFor
\State Compute the LMC mean and variance, by averaging out the estimations of each split (eq. \ref{eq:final_LMC}).

\begin{equation}
    \mathcal{M}_{N,M} = \frac{1}{S}\sum_{p=1}^S\left(\mu_{N/S, M}\right)_{p}\,,\quad\text{and}\quad \Sigma^2_{N,M} = \frac{1}{S}\sum_{p=1}^S\left(\sigma^2_{N/S, M}\right)_{p}\,.
    \label{eq:final_LMC}
\end{equation}
\end{algorithmic}
\end{algorithm}

\begin{table}[H]
    \setlength\extrarowheight{5pt}
    \centering
    \begin{tabular}{|p{0.45in}||p{1.1in}|p{1.2in}|p{1.2in}|p{1.4in}|}
        \hline
          & Monte Carlo & Machine Learning & Multilevel Monte Carlo & Lasso Monte Carlo \\
         \hline\hline
         Split & No & No & Training and evaluation sets $N_{tr} + N_{eval}=N$ & $S$ subsets of size $N/S$ \\
         \hline
         Strategy & Compute $\mu_N$ and $\sigma_N$ & Train $\widetilde f$, then evaluate $M$ samples and compute $\mu_M$ and $\sigma_M$ & Train $\widetilde f$, then compute multilevel estimators $\mu_{N_{eval}, M}$ and $\sigma_{N_{eval}, M}$ & For each subset, train $\widetilde f$, then compute  $\mu_{N/S,M}$ and $\sigma_{N/S,M}$. Average out the estimations to compute $\mathcal{M}_{N,M}$ and $\Sigma_{N,M}$\\
         \hline
         Error & Unbiased error$_{MC}\sim\frac{1}{\sqrt{N}}$ & Biased and \textit{curse of dimensionality} & Unbiased  & Unbiased error$_{LMC}\leq$error$_{MC}$\\
         \hline
    \end{tabular}
    \caption{Summary of the discussed UQ methods. Each method starts with a set of $N$ input-output samples, and $M$ input samples, with the assumption $M\gg N$.}
    \label{tab:UQ_methods}
\end{table}

\subsection{Implementation of LMC}
The LMC implementation is available in a GitLab repository \cite{noauthor_gitlab-psi_2023}. It is written in Python and makes use of standard libraries. The user can choose any surrogate model for the LMC method, but for this paper only the \texttt{Lasso} Sklearn \cite{pedregosa_scikit-learn_2011} implementation was used. The sparsity of $\beta$ in the Lasso model is controlled by a regularisation parameter $\lambda$ (see \cite{tibshirani_regression_1996}, referred to as \texttt{alpha} in the Sklearn implementations), which needs to be selected by the user. It can be chosen via model selection strategies such as cross-validation or BIC, with all strategies leading to similar results. In this paper, $\lambda$ was chosen such that the number of nonzero elements of $\beta$ would be the closest integer to $0.95\cdot N_{tr}$, for a given number of training samples $N_{tr}$.

\subsection{Simulated Assemblies}
Two representative assemblies of  UO2 and MOX fuel were used in this work, and the studied quantities of interest where the decay heat at cooling times of 2 and 50 years, and the concentrations of $^{235}$U, $^{239}$Pu, and $^{137}$Cs at EOL of the assembly. Furthermore, for the UO2 assembly, the k-effective of a canister containing the depleted fuel at EOL (fig. \ref{fig:canister}) was also computed.

The UO2 assembly is a $15\times15$ FA from the Swedish Ringhals-2 pressure water reactor (PWR), with an enrichment of $3.095\%$ $^{235}$U, and a final reported burnup of $35.7$ MWd/kgU \cite{sturek_measurements_2006}. This FA was part of a decay heat measurement campaign, and was used as a benchmark for a CASMO validation study \cite{shama_validation_2022}.

The MOX example is a $14\times14$ FA from a Swiss PWR, with an average concentration of $3.6\%$ fissile plutonium, and a final reported burnup of $58.9$ MWd/kgU \cite{primm_ariane_2002}. This FA has been analysed and compared to simulations in \cite{rochman_analysis_2021}.

\section{Results}

This section shows the results of UQ calculations with the described methods (table \ref{tab:UQ_methods}). In the following plots, and for the remainder of the paper, the UQ methods use $M=6000$ to ensure $M\gg N$. Additionally, the MLMC method uses $N_{tr}=0.8\cdot N$, and LMC uses $S=5$ (see table \ref{tab:UQ_methods} and alg. \ref{alg:LMC}).

\begin{figure}[H]
    \centering
    \includegraphics[width=\textwidth]{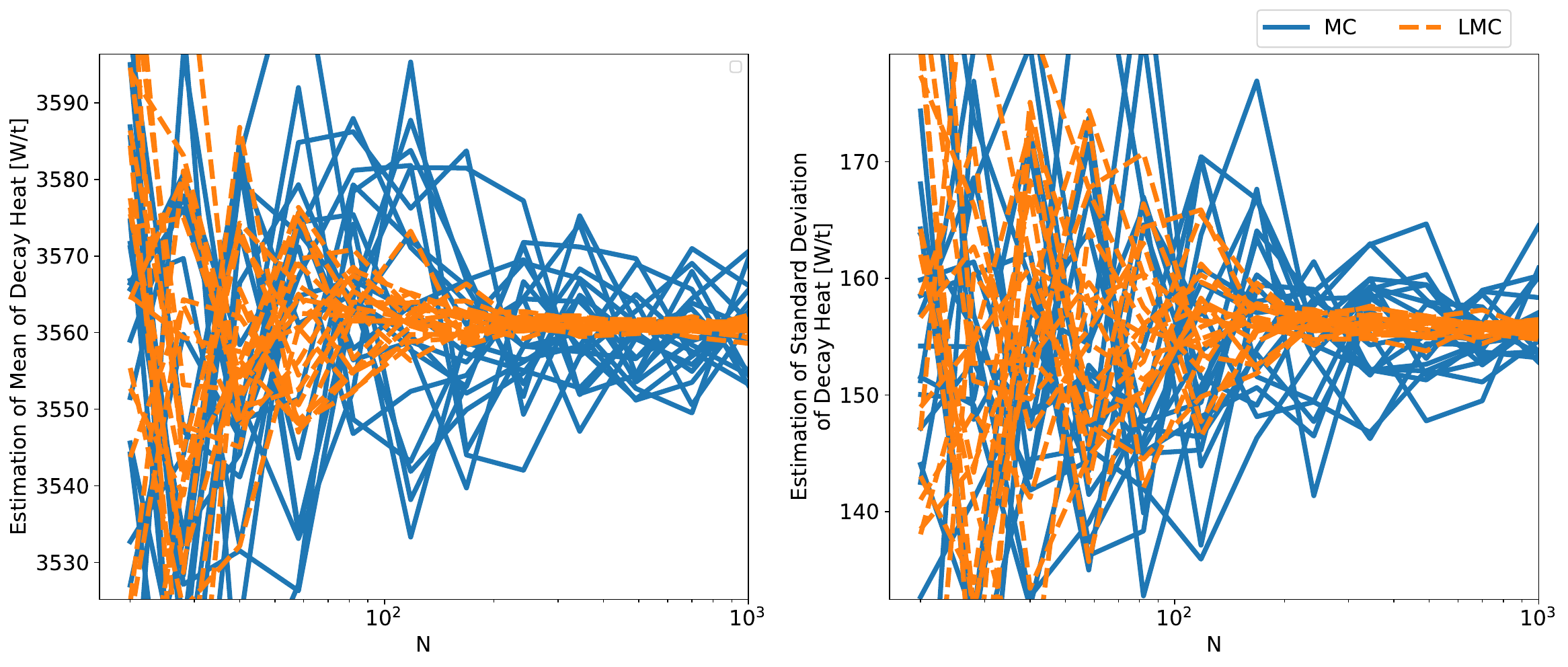}
    \caption{Estimation of mean and standard deviation of the decay heat of the UO2 assembly after 2 years of cooling. Each line is an independent UQ calculation with increasing $N$, and there are 20 lines for each method.}
    \label{fig:conv_DH}
\end{figure}

\begin{figure}[H]
    \centering
    \includegraphics[width=\textwidth, trim={0 43 0 0},clip]{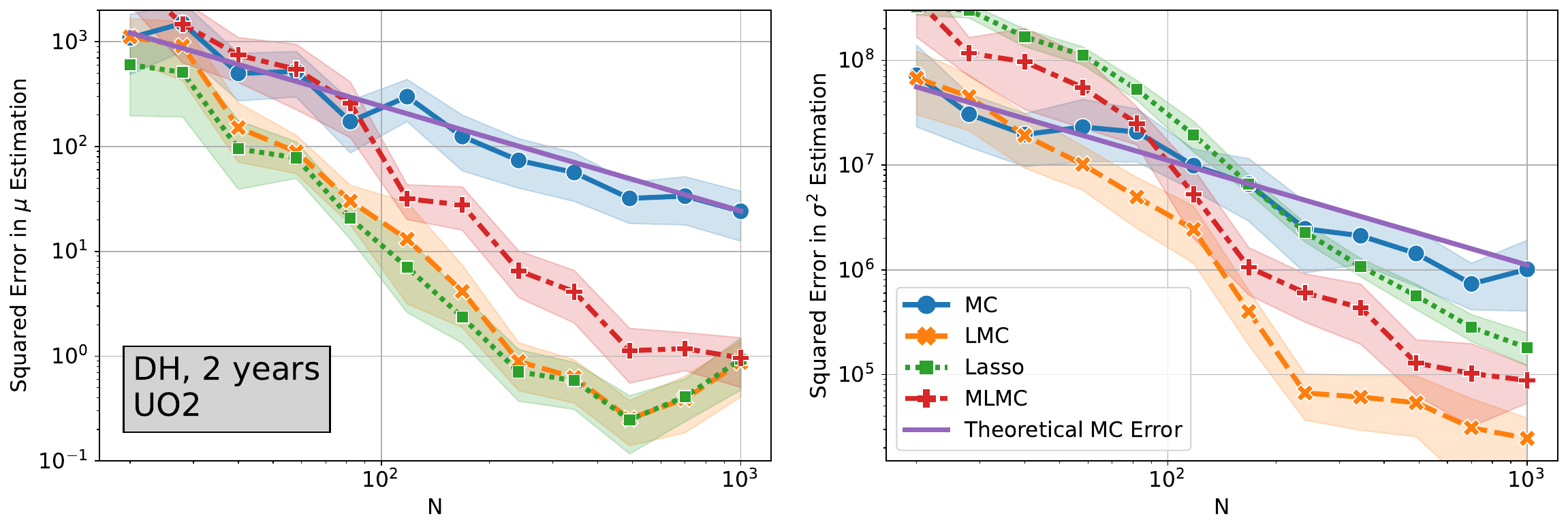}
    \includegraphics[width=\textwidth]{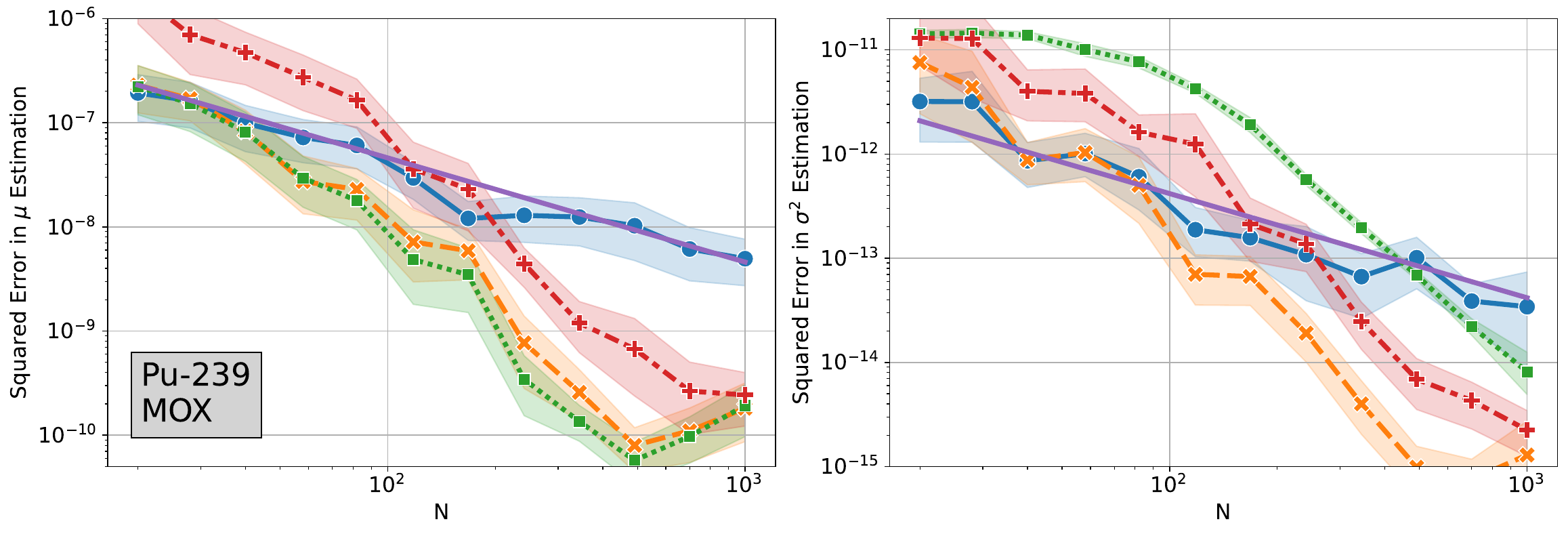}
    \caption{Convergence of the squared error, in the estimation of the mean and variance of the outputs. The studied quantities are the decay heat at 2 years of cooling for UO2, and the concentration of $^{239}Pu$ in MOX fuel at its EOL. Each line is the mean squared error of 20 independent UQ calculations, and the halos represent the $95\%$ confidence intervals.}
    \label{fig:conv_all_methods}
\end{figure}

Figure \ref{fig:conv_DH} shows 20 independent UQ calculations of the decay heat at 2 years of cooling,  with simple MC and LMC, with respect to the number of CASMO5 simulations $N$. The uncertainty is due to nuclear data perturbation of the aforementioned $15\,557$ quantities. It is clear that LMC converges faster, and that for $N>200$, the estimations become stable, whereas the simple MC method still shows large oscillations in its estimations up to $N=1000$. Only LMC and simple MC are compared in this case to allow for easier readability. A comparison between all four UQ methods can be found in figure \ref{fig:conv_all_methods}, where it can be seen that LMC is more accurate than all other methods for any $N$.

\begin{figure}[H]
    \centering
    \begin{subfigure}[b]{\textwidth}
        \centering
        \includegraphics[width=\textwidth, trim={0 110 0 0},clip]{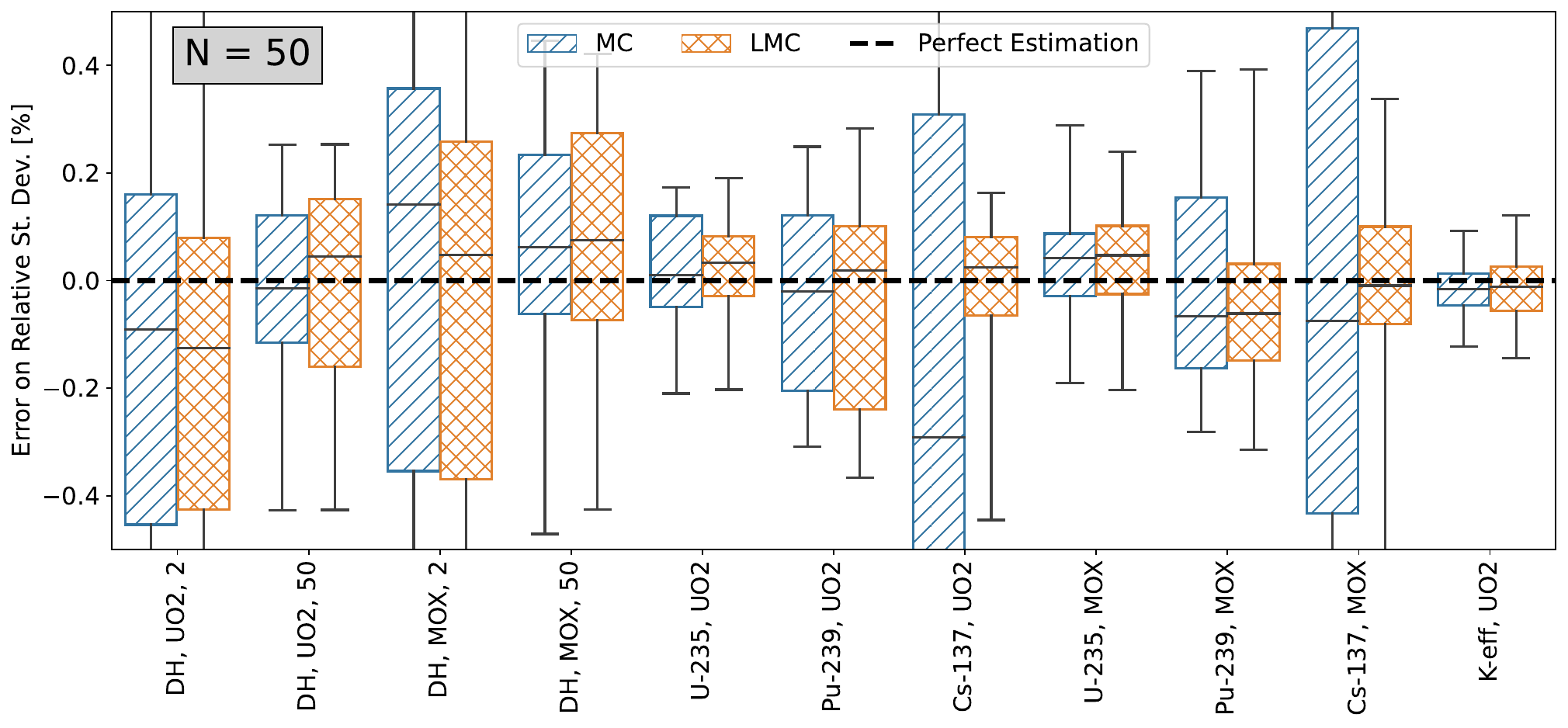}
        \includegraphics[width=\textwidth]{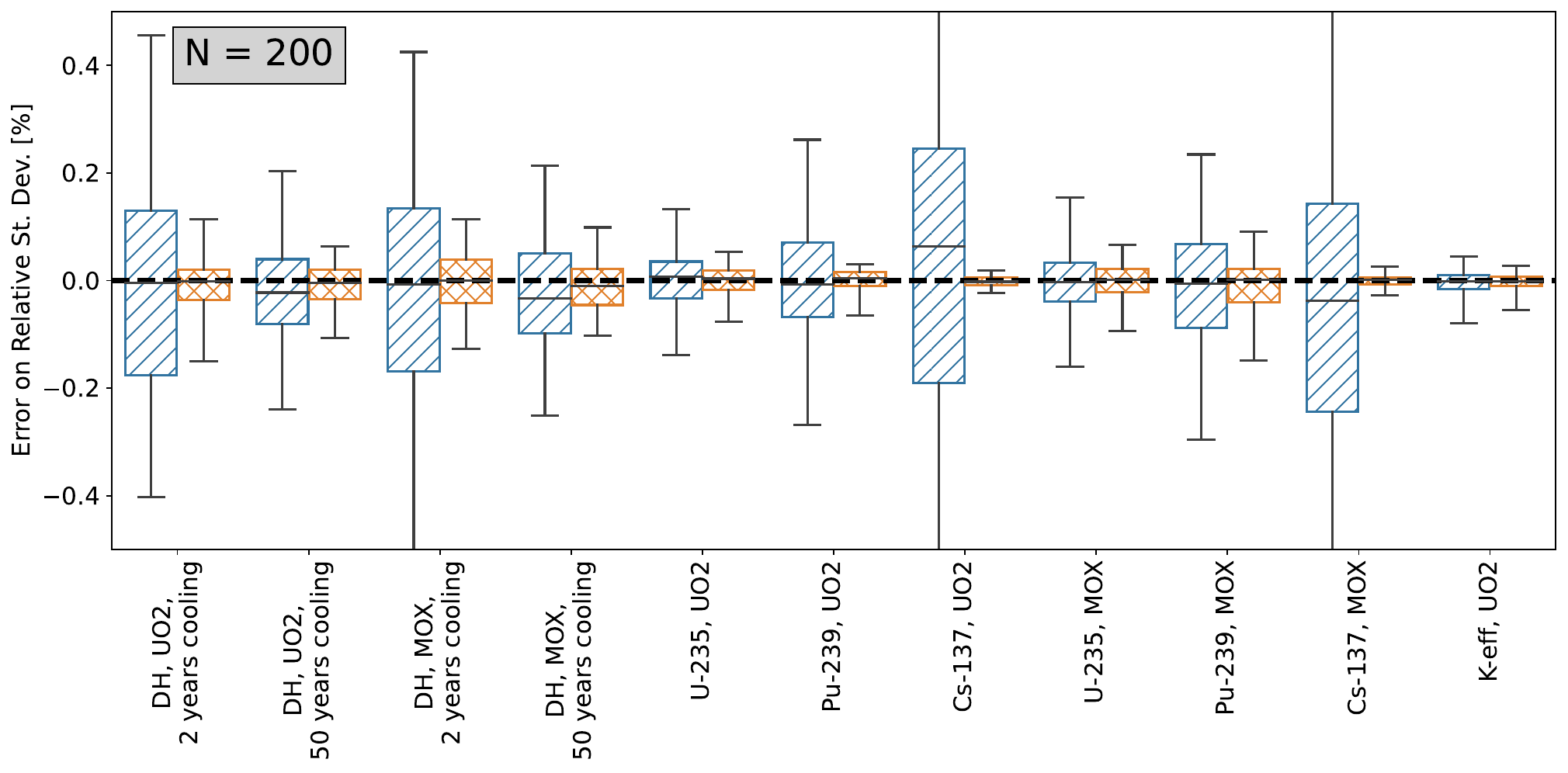}
        \caption{}
       \label{fig:boxplots}
    \end{subfigure}
    \begin{subfigure}[b]{0.45\textwidth}
        \centering
        \includegraphics[width=\textwidth]{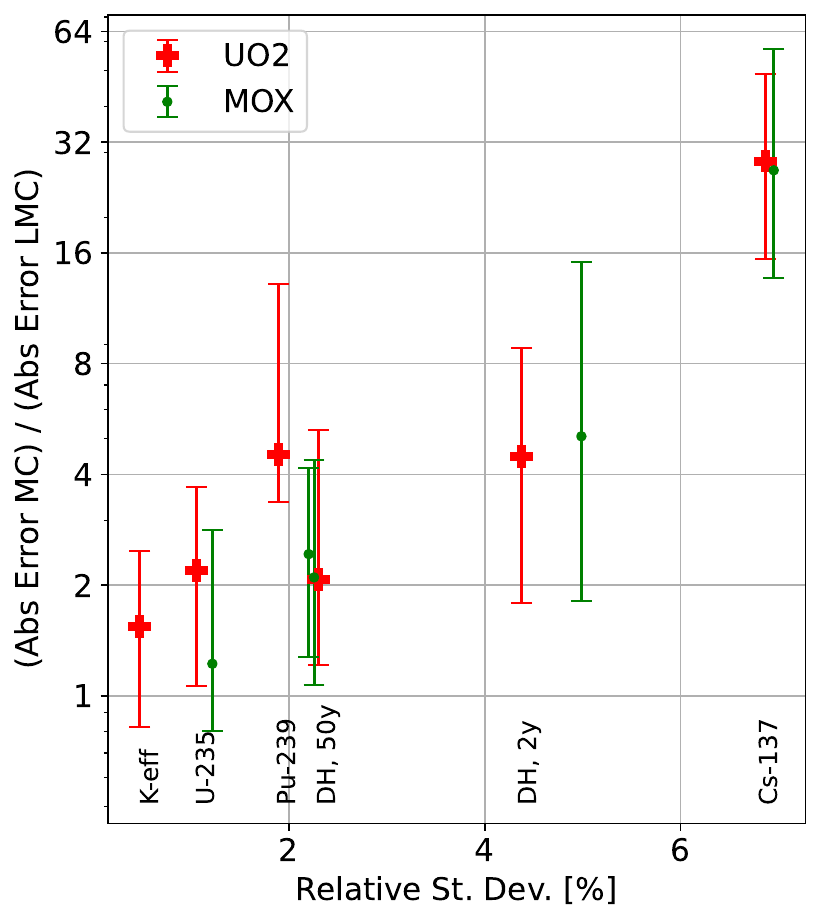}
        \caption{}
        \label{fig:improv}
    \end{subfigure}
    \begin{subfigure}[b]{0.45\textwidth}
        \centering
        \includegraphics[width=\textwidth]{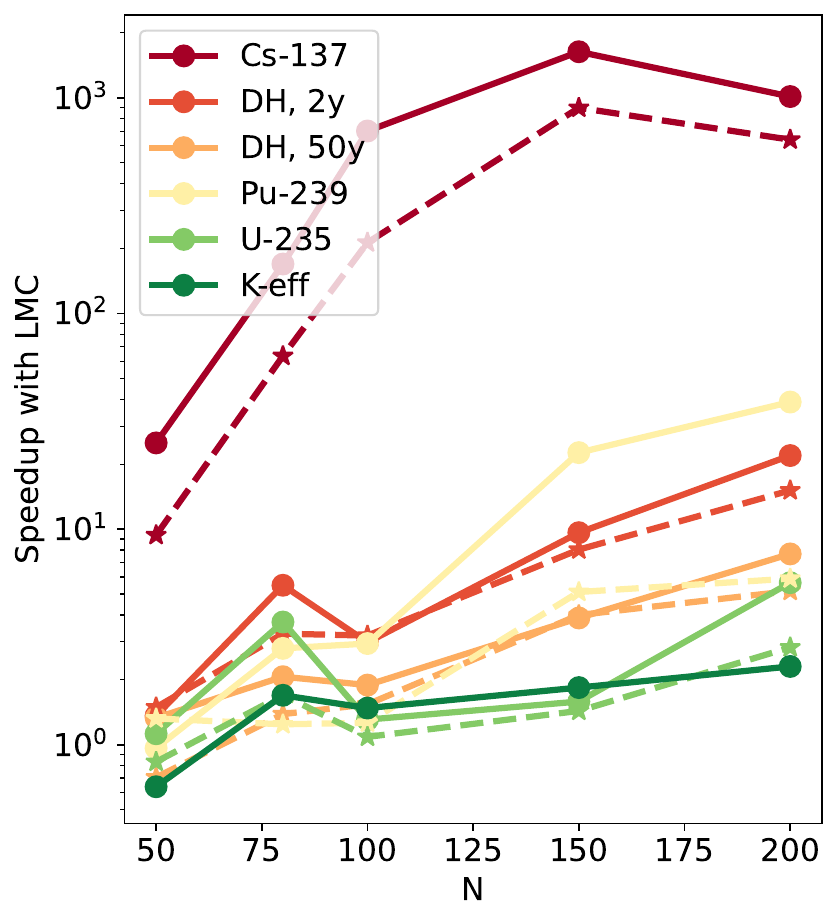}
        \caption{}
        \label{fig:speedup}
    \end{subfigure}
    \caption{(a) Errors of 20 estimations with MC and with LMC. The boxes represent the interquartile range ($50\%$ of samples fall in this range), while the whiskers represent the minimum and maximum. The line inside the boxes is the median of the errors. (b) Improvement in accuracy when using LMC versus MC, for $N=200$. The markers are the medians of 40 UQ calculations, and the errorbars are the interquantile ranges. (c) The mean speedup with LMC versus MC. Continuous lines are for the UO2 calculations, and dashed lines for the MOX ones.}
\end{figure}

Figure \ref{fig:boxplots} shows the errors with simple MC and LMC for several quantities and $N=50,200$, when estimating the relative standard deviation $100\cdot\frac{\sigma}{\mu}$. The performance of both methods varies extensively among the predicted quantities, but LMC is consistently equal or more accurate than simple MC. The quantities where MC has the largest errors are $^{137}$Cs and the decay heat at short cooling times, and both of these quantities have the largest improvement in error when using LMC. This suggests that LMC is the most advantageous when MC has large errors, and this can be further observed in figure \ref{fig:improv}. It shows the improvement in accuracy $\frac{\text{Absolute Error with MC}}{\text{Absolute Error with LMC}}$ with respect to the relative standard deviation. There appears to a be a positive correlation between these two quantities, which indicates that quantities which have the largest uncertainties due to nuclear data are the ones that benefit the most from LMC.

The computational speedup of LMC with respect to MC is plotted in figure \ref{fig:speedup} for each quantity. The speedup is measured as speedup$=\frac{N_{MC}}{N}$, where $N_{MC}$ is the average number of samples required with simple MC, to achieve the same mean squared error that LMC achieves with $N$. Similar speedups are observed between the UO2 and MOX quantities, except for the $^{239}$Pu, where the speedup is roughly 4 times smaller in the MOX case. The largest speedup is around 1000 for the concentration of $^{137}$Cs, followed by speedups in the order of 10 for the decay heats and nuclide concentrations. The lowest speedup is in the k-effective, although this could be explained by the small effect of nuclear data on this quantity -- since the nuclear data was only perturbed in the burnup calculations, the k-effective was only affected by the perturbations in the nuclide concentrations. If one were to perturb the nuclear data of the criticality calculations it is likely that the k-effective would have larger variations and hence a stronger speedup from LMC.

\section{Discussion and Conclusion}
The UQ problem of propagating the uncertainty of nuclear data is challenging due to its high-dimensionality. The best method currently in use is simple MC, which requires a large number $N$ of costly simulations. In this paper the novel method LMC \cite{alba2023lasso} was applied to this problem, as well as two other methods, multilevel Monte Carlo and Lasso machine learning. 

The LMC method is the most accurate of the four methods, and it is the only one that is equal or better than simple MC for any $N$. Although LMC is consistently more accurate than MC, its exact improvement depends largely on the studied quantity. The quantities with the largest improvement from LMC are the concentration of $^{137}$Cs and the decay heat at 2 years of cooling. Experiments suggest that quantities that show the strongest improvement from using LMC are those which are the most affected by nuclear data perturbations, i.e. those with the largest uncertainty. For the prediction of k-effective, LMC and MC performed similarly, with negligible differences. The computational speedups due to LMC are up to 1000 for predicting the concentration of $^{137}$Cs, and in the order of 10 for most other quantities. Similar results and speedups were found between the UO2 and MOX cases. 

Future work will further explore the applications of LMC to other quantities that characterise SNF. Efforts will go towards a better prediction of the k-effective uncertainty at different cooling times, and the perturbation of nuclear data will be included to criticality calculations, not only the burnup simulations.

\begin{center}
    \textbf{ACKNOWLEDGEMENTS}
\end{center}
\vspace{-11pt}
This work was part of the COLOSS and COLOSS-2 projects (COmbined Loading Optimization with Simulations and Surrogate models), funded by Swissnuclear.

\begin{center}
    \textbf{REFERENCES}
\end{center}
\vspace{-22pt}

\bibliographystyle{unsrt}
\bibliography{ref}

\begin{thebibliography}{10}

\bibitem{rochman_nuclear_2016}
D.~Rochman, A.~Vasiliev, H.~Ferroukhi, T.~Zhu, S.~C. van~der Marck, and A.~J.
  Koning.
\newblock Nuclear data uncertainty for criticality-safety: {Monte} {Carlo} vs.
  linear perturbation.
\newblock {\em Annals of Nuclear Energy}, 92:150--160, June 2016.

\bibitem{solans_loading_2020}
V.~Solans, D.~Rochman, H.~Ferroukhi, A.~Vasiliev, and A.~Pautz.
\newblock Loading optimization for {Swiss} used nuclear fuel assemblies into
  final disposal canisters.
\newblock {\em Nuclear Engineering and Design}, 370:110897, December 2020.

\bibitem{ebiwonjumi_machine_2021}
B.~Ebiwonjumi, A.~Cherezov, S.~Dzianisau, and D.~Lee.
\newblock Machine learning of {LWR} spent nuclear fuel assembly decay heat
  measurements.
\newblock {\em Nuclear Engineering and Technology}, 53(11):3563--3579, November
  2021.

\bibitem{beyer_when_1999}
K.~Beyer, J.~Goldstein, R.~Ramakrishnan, and U.~Shaft.
\newblock When {Is} “{Nearest} {Neighbor}” {Meaningful}?
\newblock In Catriel Beeri and Peter Buneman, editors, {\em Database {Theory}
  — {ICDT}’99}, Lecture {Notes} in {Computer} {Science}, pages 217--235,
  Berlin, Heidelberg, 1999.

\bibitem{jolliffe_principal_2016}
I.~T. Jolliffe and J.~Cadima.
\newblock Principal component analysis: a review and recent developments.
\newblock {\em Philosophical Transactions of the Royal Society A: Mathematical,
  Physical and Engineering Sciences}, 374(2065):20150202, April 2016.
\newblock Publisher: Royal Society.

\bibitem{subspace_2014}
P.~G. Constantine, E.~Dow, and Q.~Wang.
\newblock Active subspace methods in theory and practice: Applications to
  kriging surfaces.
\newblock {\em SIAM Journal on Scientific Computing}, 36(4):A1500--A1524, 2014.

\bibitem{alba2023lasso}
A.~Albà, R.~Boiger, D.~Rochman, and A.~Adelmann.
\newblock {Lasso Monte Carlo, a Variation on Multi Fidelity Methods for High
  Dimensional Uncertainty Quantification}.
\newblock 2023.
\newblock arXiv:2210.03634.

\bibitem{tibshirani_regression_1996}
R.~Tibshirani.
\newblock Regression {Shrinkage} and {Selection} via the {Lasso}.
\newblock {\em Journal of the Royal Statistical Society. Series B
  (Methodological)}, 58(1):267--288, 1996.

\bibitem{giles_multilevel_2008}
M.~B. Giles.
\newblock Multilevel {Monte} {Carlo} {Path} {Simulation}.
\newblock {\em Operations Research}, 56(3):607--617, June 2008.

\bibitem{rhodes_casmo-5_2006}
J.~Rhodes, K.~Smith, and D.~Lee.
\newblock {CASMO}-5 development and applications.
\newblock Technical report, American Nuclear Society - ANS; La Grange Park
  (United States), July 2006.

\bibitem{shama_validation_2022}
A.~Shama, D.~Rochman, S.~Caruso, and A.~Pautz.
\newblock Validation of spent nuclear fuel decay heat calculations using
  {Polaris}, {ORIGEN} and {CASMO5}.
\newblock {\em Annals of Nuclear Energy}, 165:108758, January 2022.

\bibitem{wieselquist_psi_2013}
W.~Wieselquist, T.~Zhu, A.~Vasiliev, and H.~Ferroukhi.
\newblock {PSI} {Methodologies} for {Nuclear} {Data} {Uncertainty}
  {Propagation} with {CASMO}-{5M} and {MCNPX}: {Results} for {OECD}/{NEA} {UAM}
  {Benchmark} {Phase} {I}.
\newblock {\em Science and Technology of Nuclear Installations}, 2013:1--15,
  2013.

\bibitem{aures_benchmarking_2017}
A.~Aures, F.~Bostelmann, M.~Hursin, and O.~Leray.
\newblock Benchmarking and application of the state-of-the-art uncertainty
  analysis methods {XSUSA} and {SHARK}-{X}.
\newblock {\em Annals of Nuclear Energy}, 101:262--269, March 2017.

\bibitem{chadwick_endfb-vii1_2011}
M.~B. Chadwick, M.~Herman, P.~Obložinský, M.~E. Dunn, Y.~Danon, A.~C. Kahler,
  D.~L. Smith, B.~Pritychenko, G.~Arbanas, R.~Arcilla, R.~Brewer, D.~A. Brown,
  R.~Capote, A.~D. Carlson, Y.~S. Cho, H.~Derrien, K.~Guber, G.~M. Hale,
  S.~Hoblit, S.~Holloway, T.~D. Johnson, T.~Kawano, B.~C. Kiedrowski, H.~Kim,
  S.~Kunieda, N.~M. Larson, L.~Leal, J.~P. Lestone, R.~C. Little, E.~A.
  McCutchan, R.~E. MacFarlane, M.~MacInnes, C.~M. Mattoon, R.~D. McKnight,
  S.~F. Mughabghab, G.~P.~A. Nobre, G.~Palmiotti, A.~Palumbo, M.~T. Pigni,
  V.~G. Pronyaev, R.~O. Sayer, A.~A. Sonzogni, N.~C. Summers, P.~Talou, I.~J.
  Thompson, A.~Trkov, R.~L. Vogt, S.~C. van~der Marck, A.~Wallner, M.~C. White,
  D.~Wiarda, and P.~G. Young.
\newblock {ENDF}/{B}-{VII}.1 {Nuclear} {Data} for {Science} and {Technology}:
  {Cross} {Sections}, {Covariances}, {Fission} {Product} {Yields} and {Decay}
  {Data}.
\newblock {\em Nuclear Data Sheets}, 112(12):2887--2996, December 2011.

\bibitem{romano_openmc_2015}
P.~K. Romano, N.~E. Horelik, B.~R. Herman, A.~G. Nelson, B.~Forget, and
  K.~Smith.
\newblock {OpenMC}: {A} state-of-the-art {Monte} {Carlo} code for research and
  development.
\newblock {\em Annals of Nuclear Energy}, 82:90--97, August 2015.

\bibitem{vasiliev_preliminary_2019}
A.~Vasiliev, J.~Herrero, M.~Pecchia, D.~Rochman, H.~Ferroukhi, and S.~Caruso.
\newblock Preliminary {Assessment} of {Criticality} {Safety} {Constraints} for
  {Swiss} {Spent} {Nuclear} {Fuel} {Loading} in {Disposal} {Canisters}.
\newblock {\em Materials}, 12(3):494, January 2019.

\bibitem{krumscheid_quantifying_2020}
S.~Krumscheid, F.~Nobile, and M.~Pisaroni.
\newblock Quantifying uncertain system outputs via the multilevel {Monte}
  {Carlo} method — {Part} {I}: {Central} moment estimation.
\newblock {\em Journal of Computational Physics}, 414:109466, August 2020.

\bibitem{noauthor_gitlab-psi_2023}
{GitLab}-{PSI}: {LassoMonteCarlo}.
\newblock \url{https://gitlab.psi.ch/albajacas_a/lassomontecarlo}, May 2023.

\bibitem{pedregosa_scikit-learn_2011}
F.~Pedregosa, G.~Varoquaux, A.~Gramfort, V.~Michel, B.~Thirion, O.~Grisel,
  M.~Blondel, P.~Prettenhofer, R.~Weiss, V.~Dubourg, J.~Vanderplas, A.~Passos,
  D.~Cournapeau, M.~Brucher, M.~Perrot, and É. Duchesnay.
\newblock Scikit-learn: {Machine} {Learning} in {Python}.
\newblock {\em Journal of Machine Learning Research}, 12(85):2825--2830, 2011.

\bibitem{sturek_measurements_2006}
F.~Sturek, L.~Agrenius, and O.~Osifo.
\newblock Measurements of decay heat in spent nuclear fuel at the {Swedish}
  interim storage facility, {Clab}.
\newblock Technical Report R-05-62, Svensk Kärnbränslehantering AB, December
  2006.

\bibitem{primm_ariane_2002}
RT~Primm.
\newblock {ARIANE} {International} {Programme} {Final} {Report}.
\newblock Technical report, United States, May 2002.
\newblock ORNL/SUB--97-XSV750-1 INIS Reference Number: 34070003.

\bibitem{rochman_analysis_2021}
D.~Rochman, A.~Vasiliev, H.~Ferroukhi, and M.~Hursin.
\newblock Analysis for the {ARIANE} {BM1} and {BM3} samples: nuclide inventory
  and decay heat.
\newblock {\em EPJ Nuclear Sciences \& Technologies}, 7:18, 2021.

\end{thebibliography}

\end{document}